\documentclass[a4paper,twocolumn,preprint]{emulateapj}
\usepackage{graphicx}
\slugcomment{Accepted for publication in ApJ Letters}

\shorttitle{Long term evolution of $\nu$ cooled disks and flares in GRBs}

\shortauthors{Lee, Ramirez-Ruiz \& L\'{o}pez-C\'{a}mara}

\begin{document}

\title{Phase transitions and He-synthesis driven winds in neutrino
  cooled accretion disks: prospects for late flares in short gamma-ray
  bursts}

\author{William H. Lee\altaffilmark{1}, Enrico
  Ramirez-Ruiz\altaffilmark{2} and Diego
  L\'{o}pez-C\'{a}mara\altaffilmark{1}}\altaffiltext{1}{Instituto de
  Astronom\'{\i}a, Universidad Nacional Aut\'{o}noma de M\'{e}xico,
  Apdo. Postal 70-264, M\'{e}xico D.F. 04510,
  MEXICO}\altaffiltext{2}{Department of Astronomy and Astrophysics,
  University of California, Santa Cruz, CA 95064, USA}

\begin{abstract}
We consider the long term evolution of debris following the tidal
disruption of compact stars in the context of short gamma ray bursts
(SGRBs). The initial encounter impulsively creates a hot, dense,
neutrino-cooled disk capable of powering the prompt emission.  After a
long delay, we find that powerful winds are launched from the surface
of the disk, driven by the recombination of free nucleons into
$\alpha$-particles. The associated energy release depletes the mass
supply and eventually shuts off activity of the central engine. As a
result, the luminosity and mass accretion rate deviate from the
earlier self-similar behavior expected for an isolated ring with
efficient cooling. This then enables a secondary episode of delayed
activity to become prominent as an observable signature, when material
in the tidal tails produced by the initial encounter returns to the
vicinity of the central object. The time scale of the new accretion
event can reach tens of seconds to minutes, depending on the details
of the system. The associated energies and time scales are consistent
with those occurring in X-ray flares.
\end{abstract}


\section{Introduction}\label{sec:intro}

Gamma Ray Bursts (GRBs) probably involve newborn compact objects
accreting at rates high enough that neutrinos are the dominant source
of cooling. Short bursts (SGRBs) exhibit a substantial degree of
diversity \citep{nakar07}, although a large fraction of them probably
come from an old stellar progenitor population \citep{lrr07}, while
long events (LGRBs) have been associated to the collapse of massive
stars \citep{wb06}. Although two classes are probably present, some
overlap occurs and making a distinction for individual events is not
entirely straightforward \citep{zhang09}. Since the launch of the {\it
  Swift} satellite it has been possible to study the onset of the
afterglow emission, thought to be produced when the inertia of the
swept up external matter starts to slow down the ejecta appreciably
\citep{fm06,obrien06,gcn07}.  This has revealed in many cases renewed
rapid flaring (usually in soft X-rays) long after the canonical prompt
gamma-ray emission has ceased -- at timescales that are much longer
than the usual dynamical or even viscous time scales considered for
the accretion process itself. In SGRBs these flares typically appear
30-60~s after the burst and last for tens of seconds as well,
containing a non-negligible fraction of the radiated energy. There is
a general consensus that they must involve renewed activity within the
central engine, though simple broad pulses could arise from an
external shock interaction \citep{rr01,zhang06,nousek06}.

We seek to accurately compute the evolution of these neutrino cooled
structures on long time scales, and consider mechanisms which might
allow for the production of a new episode of energy release. Disk
fragmentation \citep{perna06}, fall back tidal tails
\citep{rosswog07a}, strong magnetic fields in rapidly rotating pulsars
\citep{kr98,rrr02,dai06}, magnetic halting \citep{proga06} and cooling
transitions \citep{lazzati08} , among others, may be related to such
secondary transients, but no definitive answers have been
forthcoming. In this {\it Letter} we study the thermodynamical
properties of these flows in detail in an attempt to determine how
they may be relevant to their long term evolution, which we compute
numerically.

\section{Input physics and numerical study}\label{sec:input}

Previous work on $\nu$--cooled disks has shown the importance of the
proper consideration of thermodynamics
\citep{woosley93,mw99,pwf99,npk01,dmpn02,km02,bel03,srj04,lrrp05,rosswog07b,cb07}. The
densities and temperatures are typically $\rho \simeq
10^{10}$g~cm$^{-3}$ and $T \simeq 10^{10}$~K, and we consider an ideal
gas of free nucleons and $\alpha$ particles in nuclear statistical
equilibrium (NSE), black-body radiation, relativistic e$^{\pm}$ pairs
of arbitrary degeneracy, and neutrinos, for which we use a simplified
two-stream approximation \citep{lclrr09}. Neutronization in the
optically thick and thin regimes is considered by detailed balance of
weak interactions \citep{lrrp05}, which consistently accounts for the
optically thick and thin regimes. Emissivities are taken from the
fitting functions given by \citet{itoh96} for pair annihilation and
\citet{lmp01} for e$^{\pm}$ capture onto nucleons, and the energy from
photodisintegration of $\alpha$ particles is also included in the
energy equation.

The evolution of the disk is followed with a two dimensional
Lagrangian Smooth Particle Hydrodynamics (SPH) code \citep{mon92} in
azimuthal symmetry. The central mass produces a Paczynski-Wiita
\citep{pw80} pseudo-Newtonian potential, reproducing the position of
the last stable orbit for a Swarzschild black hole (BH), and accretion
is implemented with an absorbing boundary at $r_{\rm in}=2 r_{\rm
  Sch}=4GM_{\rm BH}/c^{2}$. We also include a boundary at $r_{\rm
  out}\simeq 200 r_{\rm Sch}$, where outflowing fluid elements are
removed from the calculation, and use an $\alpha$-prescription
\citep{ss73} for the magnitude of the viscosity, with $10^{-3} \leq
\alpha \leq 10^{-1}$. To avoid catastrophic loss of resolution a
fission routine maintains a predetermined minimum number of fluid
elements throughout the evolution. The initial disk mass is $0.003
\leq M_{\rm disk}/M_{\odot} \leq 0.3$, and the initial BH mass is
$M_{\rm BH}\simeq 5M_{\odot}$.

\section{Isolated disk evolution and the importance of He synthesis}\label{sec:winds}

A ring in centrifugal equilibrium at characteristic radius $R$
accretes and evolves on a viscous time scale if angular momentum is
transported outwards and mass inwards. The standard approach is to
consider that viscous stresses, and the associated energy dissipation
can be parameterized through the $\alpha$ prescription, with the
coefficient of viscosity given by $\nu=\alpha c_s^{2}/\Omega_{\rm
  Kep}$, where $c_{\rm s}$ and $\Omega_{\rm Kep}$ are the local sound
speed and Keplerian orbital frequency, respectively. The ring spreads
radially on a timescale $t_{\rm visc}\simeq R/10 \nu$, eventually
transferring all of the angular momentum to an infinitesimal amount of
mass at infinity. If the dissipated energy is radiated efficiently it
remains geometrically thin, with aspect ratio $h/r \simeq c_{\rm
  s}/v_{\phi} \ll 1$. In the context of GRBs, where accretion rates
can reach $\simeq 1$~$M_{\odot}$ ~s$^{-1}$, the disk cools by neutrino
emission and typically the accretion efficiency is $\eta_{\rm acc}
=L_{\nu}/ \dot{M}c^{2} \simeq 0.01-0.1$.  The nature of the mechanism
is quite irrelevant in determining the vertical structure of the disk,
so long as it is minimally proficient in removing internal energy
\citep{lrr02}. In this case the decay in luminosity and accretion rate
are closely correlated.

An earlier study considering accretion disks formed impulsively by the
tidal disruption of main sequence stars by supermassive BHs by
\citet{cannizzo90} found that the luminosity follows a power law in
time, $L \propto t^{-\beta}$, with $\beta \simeq 1.2$, as does
$\dot{M}$. A key consideration is that the mass of the disk be only
removed through accretion onto the central object.  Recently,
\citet{metzger08} have carefully computed the evolution of a ring of
matter in the $\nu$--cooled regime under similar assumptions as the
study of \citet{cannizzo90}. The evolution displays similar behavior,
and they report a decay index $\beta \simeq 4/3$. Further, they
consider generic solutions in which a disk-driven wind is present,
extracting mass as well as angular momentum from the fluid. In that
case both $L$ and $ \dot{M}$ enter a phase of rapid decline once mass
driving through the wind becomes important, typically after a fraction
of a second.

\begin{figure}
  \begin{center}
    \includegraphics[height=\textheight]{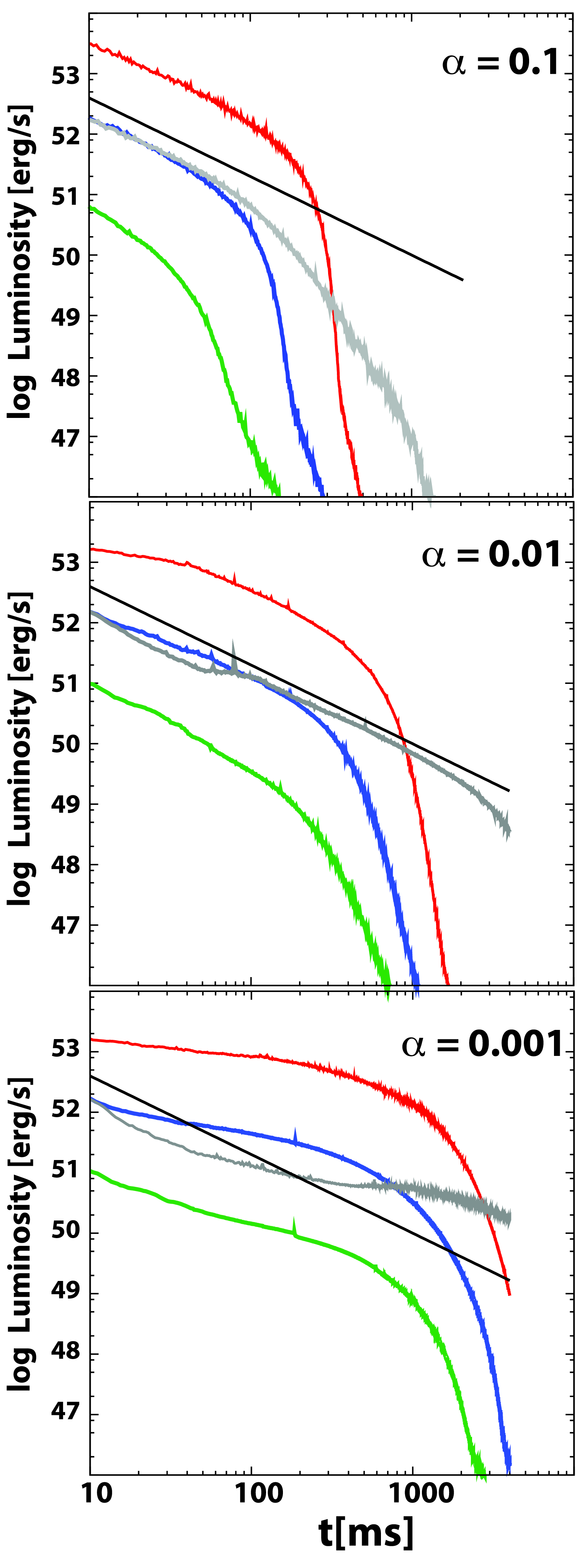}
    \caption{Neutrino luminosity for models with $M_{\rm
        disk}/M_{\odot}=0.3,0.03,0.003$ (top to bottom in each panel)
      and $\alpha=0.1,0.01,0.001$ (top to bottom panels). The grey
      lines show the result of calculations in which the energy
      associated with He synthesis was not taken into account, and
      exhibit a characteristic power law decay with index $\beta
      \simeq 1.3$, shown for reference.}
    \label{fig:lightcurves}
  \end{center}
\end{figure} 

Thus motivated, we have computed the long term evolution of
$\nu$-cooled accretion disks under the assumptions given in
\S~\ref{sec:input}. The early stages ($t \leq 0.5$~s) are
characterized as shown before by the release of the initial reservoir
of internal energy left over from the merger phase, and are followed
by a more gradual decay in which angular momentum transport and energy
dissipation contribute to both draining the disk of matter and
powering the neutrino luminosity, $L_{\nu}$. Depending on the disk's
mass, the flow can be optically thick to neutrinos in the early phases
and this can have an effect on the global energy release.

After a longer delay, powerful winds are launched from the disk
because of He synthesis from the available free nucleons, at the
characteristic radius where the nuclear binding energy of He is
roughly equal to the gravitational binding energy, $GM_{\rm BH}m_{\rm
  p}/r_{\rm wind}\simeq 7$~MeV. When the available mass approaches
$r_{\rm wind}$, it is driven away almost isotropically, rapidly
depleting the disk and shutting down neutrino emission. The effects
are seen in Figure~\ref{fig:lightcurves} where $L_{\nu}$ is plotted
for simulations covering two orders of magnitude in initial disk mass
and strength of viscous transport. For comparison, curves of benchmark
simulations in which the energy associated with He synthesis,
7~MeV/baryon, was {\em not} taken into account in the energy equation
are also shown (grey lines), and clearly follow a power law with index
$\beta \simeq 1.3$ for a longer time, close to that predicted by
\citet{metzger08} (see especially the case with $\alpha=0.01$). The
break to something more akin to exponential decay is entirely due to
the driving of winds off the surface of the disk. The mass flow rates
and mechanical power are $(\dot{M}
\mbox{[$10^{-2}~M_{\odot}$~s$^{-1}$]},L_{\rm
  wind}\mbox{[$10^{50}$~erg~s$^{-1}$]})=(1.5,1.3), (0.2,0.15),
(0.025,0.01)$ for initial disk mass $M_{\rm disk}=0.3, 0.03, 0.003
M_{\odot}$, respectively, and $\alpha=0.01$.

In the reference simulations which do not consider the nuclear binding
energy, $L_{\nu}$ eventually also deviates from the simple power law
because the mass drains into the BH as a result of viscous transport
and neutrino cooling shuts off. As expected, the energy normalization
is $L_{\nu} \propto M_{\rm disk}$, since the energy reservoir is
gravitational in nature, and the temporal breaks scale with the vigor
of angular momentum transport since a lower $\alpha$ implies a longer
delay in the mass moving to $r_{\rm wind}$.

\begin{figure}
  \begin{center}
    \includegraphics[width=1.0\columnwidth]{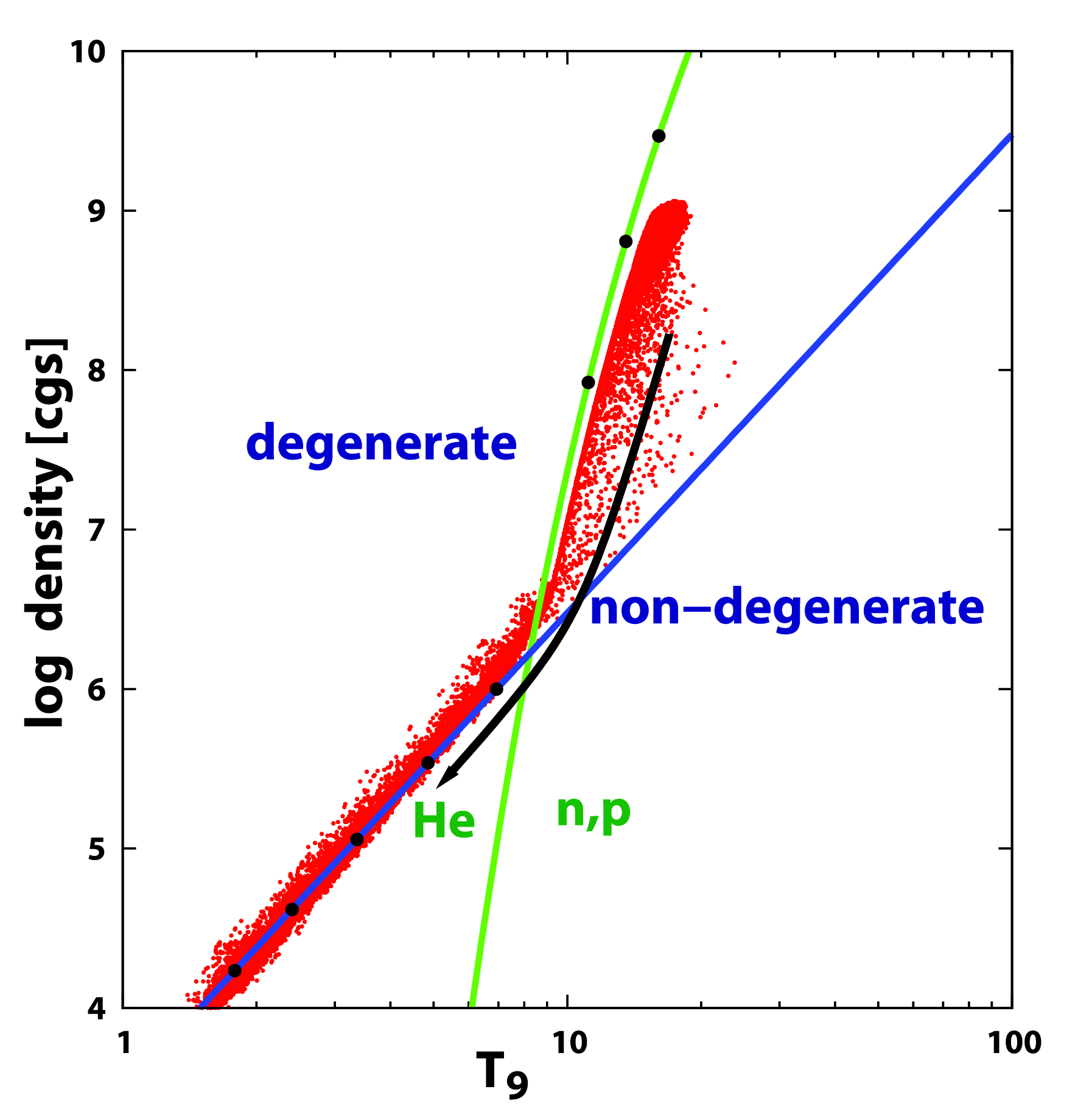}
    \caption{Time evolution of a neutrino cooled accretion disk in the
      density--temperature plane. The blue power law marks a line of
      constant degeneracy given by $kT \propto \rho^{1/3}$.  The
      curved green line shows the transition in composition from free
      nucleons at high temperatures and low densities to $\alpha$
      particles in the opposite extreme assuming NSE.
      The thick black arrow marks the flow of the gas as
      the disk is drained through winds and accretion onto
      the central mass. The spacing
      between each successive pair of black dots along the blue and
      green curves corresponds to a change in the mass cooling rate
      (in erg~g$^{-1}$~s$^{-1}$) of one order of magnitude. The
      highest point has the largest value at
      $\dot{q}_{\nu}=10^{20}$erg~g$^{-1}$~s$^{-1}$. The red dots show
      the values of the density and temperature in the disk for a
      snapshot in the evolution of the model with $\alpha=0.01$ and
      initial disk mass $M_{\rm disk}=0.03M_{\odot}$.}
    \label{fig:rhoT}
  \end{center}
\end{figure}

The thermodynamical conditions clearly affect the evolution of the
accretion disk. Figure~\ref{fig:rhoT} shows a snapshot in the
evolution of one of our models. The two main cooling mechanisms are,
as mentioned, e$^{\pm}$ annihilation and e$^{\pm}$ capture onto free
nucleons. The first requires abundant e$^{\pm}$ pair creation, while
the second only operates if there are free nucleons in the fluid. As
the transition from $\alpha$ particles to a free gas of neutrons and
protons is quite rapid, and the abundance of pairs decreases
exponentially with degeneracy, these two conditions clearly mark the
boundaries where the gas is allowed to lie in such a diagram in order
to cool effectively. If a fluid element were to enter a region where
cooling is inoperative, the associated expansion will quickly cause a
drop in density and force it back into the cooling region (pressure
mainly comes from the ideal gas terms in the equation of state). As
the disk is depleted, it follows the photodisintegration line closely
until this crosses the degeneracy threshold, with $T \propto
\rho^{1/3}$, and then continues to even lower densities along parallel
tracks. It is this transition to a disk dominated in composition by
$\alpha$ particles which leads to the production of strong winds and
rapidly exhausts its mass as discussed above. The change occurs at
$\log \rho[\mbox{g~cm$^{-3}$}] \simeq 6.5$ and $T\simeq 10^{10}$~K.

\section{The effects of mass injection at late times}\label{sec:injection}

The tidal disruption of a neutron star by a more massive BH, or a
double neutron star merger, ejects stellar material into long tidal
tails through the outer Lagrange point
\citep{ls74,rs94,lee01,rosswog05,rosswog07a}. A fraction of this is
has enough energy to escape the system, with possibly interesting
nucleosynthetic consequences. The rest is bound to the central object
on highly eccentric orbits and a range of initial
energies. Trajectories are essentially ballistic and this can provide
an estimate for the rate of mass fall back. A constant differential
mass distribution with energy yields homologous flow and $\dot{M}_{\rm
  fb} =\dot{M}_0 (t/t_0)^{-5/3}$, where $t_0$ and $\dot{M}_{0}$ fix
the normalization, and the particular details depend on the equation
of state of nuclear matter and the initial mass ratio. Simulations
indicate that up to $10^{-1} M_{\odot}$ of material may follow this
behavior. The bulk of the mass returns to small radii after a few
seconds, and the question is whether it can release its gravitational
binding energy efficiently and at the same time alter the overall
evolution of the accretion flow.

We now proceed to describe the evolution of the fall back material
based on our knowledge of merger event properties and the resulting
configurations.  At some time $t_{\rm inject}\simeq 1-10$~s, a mass
$M_{\rm fb}\simeq 10^{-2}$~M$_{\odot}$ creates a new ring with
characteristic radius $R_{\rm } \simeq 3-5 \times 10^{7}$~cm. This is
located further out than the original accretion torus, typically by a
factor of 2-4 because of its somewhat larger specific angular momentum
after ejection in the tidal disruption of the neutron star. Note that
the fall back mass accretion rate is only that at which the tidal
tails feed the disk, and {\em not} the rate at which the central
object grows. The new ring will dominate the evolution of the system
provided that the injected mass is greater than that remaining in the
original disk at $t=t_{\rm inject}$, $M_{\rm fb}(t_{\rm inject}) \gg
M_{\rm disk}(t_{\rm inject})$.  For an integrated value of $M_{\rm fb}
\simeq 5 \times 10^{-2}$~M$_{\odot}$ we find that this is valid at
times ranging from 0.4-3~s, depending on the initial value of
$\alpha$. The key ingredients are the specific angular momentum and
total mass of the injected material, and the strength of viscous
transport.  Not all combinations can lead to a clearly defined
secondary episode of energy release (e.g., if the re-injection is too
close to the BH and/or too slow, or lacks sufficient mass, the effect
will be greatly reduced).

\begin{figure}
  \begin{center}
    \includegraphics[width=1.0\columnwidth]{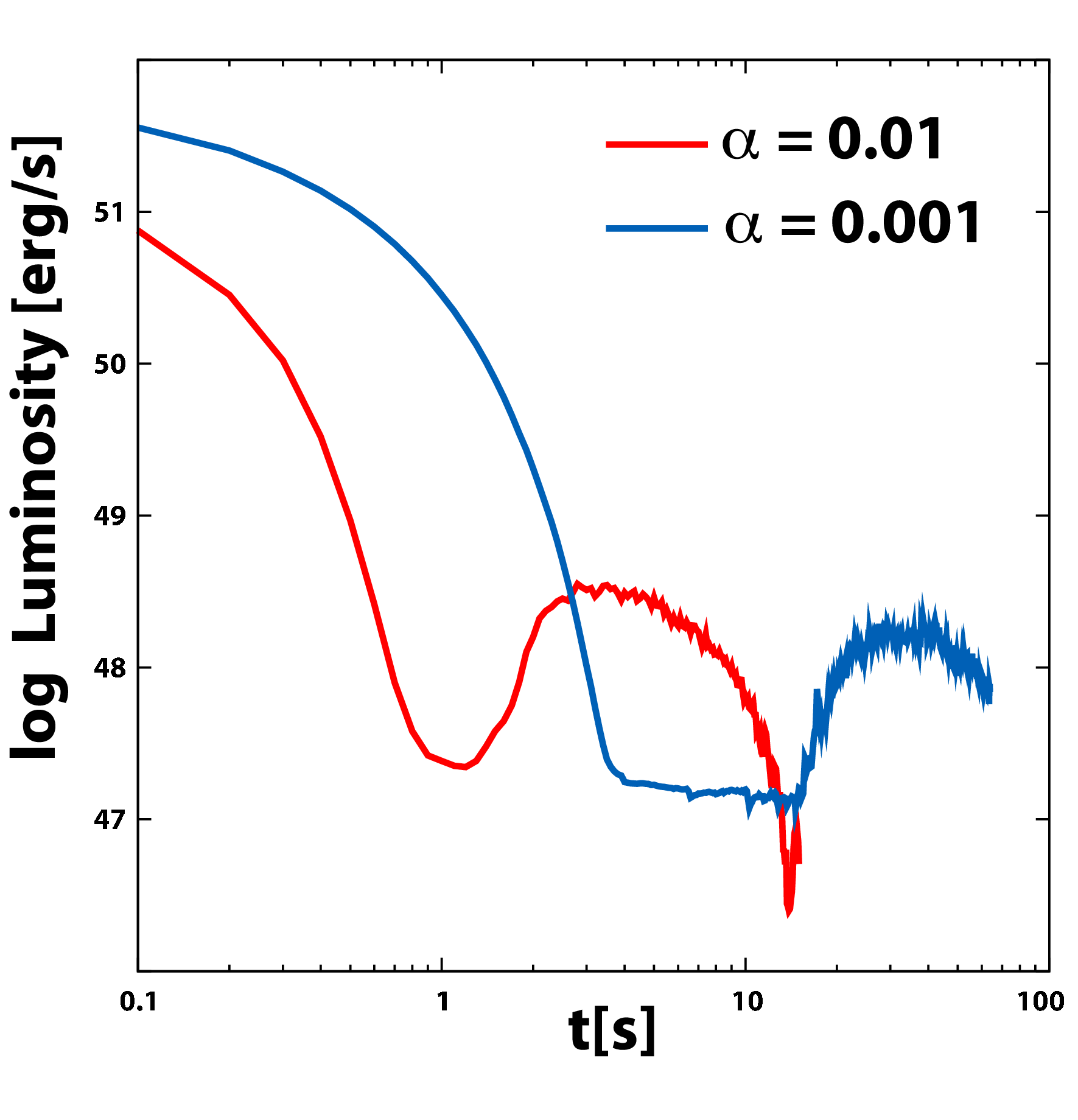}
    \caption{Neutrino luminosity for the evolution of post merger
      disks and including the contribution from fall back with initial
      mass $M_{\rm disk}/M_{\odot}=0.03$ and $M_{\rm
        fb}/M_{\odot}=0.02$. The energy normalization scales linearly
      with the mass. Note that in this plot the time is given in
      seconds. The red and blue curves show the evolution for
      $\alpha=0.01$ and $\alpha=0.001$, respectively, where the
      initial decay is given by the original disk and the late
      emission peaking at $t\simeq 3,30$~s comes from the material
      injected at $t_{\rm inject}\simeq 1$~s. The time delay until the
      flare maximum is clearly dependent on the intensity of angular
      momentum transport.}
    \label{fig:ringlight}
  \end{center}
\end{figure}

We show the evolution of rings with initial masses $M_{\rm fb}=0.2,
0.02$~M$_{\odot}$ and viscosity $\alpha=0.01,0.001$, which produce
flare-like activity. As the ring spreads radially, the temperature and
density in the innermost regions rise on a time scale that depends on
$\alpha$, and as a result the luminosity increases. The fluid is now
mostly $\alpha$ particles and the main source of cooling is e$^{\pm}$
annihilation, which is efficient enough to avoid unbinding the gas and
depleting the disk's mass. The accretion rate and luminosity
subsequently drop off as a power law with index $\beta \simeq 1.3$, as
in the previous simplified calculations where no winds were allowed to
form, before making a transition to a steeper decay as neutrino
cooling becomes less efficient (see Figure~\ref{fig:ringlight}). The
peak neutrino luminosity reaches
$\log[L_{\nu}(\mbox{erg~s$^{-1}$})]\simeq 48-49$, and can remain above
$\simeq 10^{47}$~erg~s$^{-1}$ for up to $t\simeq 100$~s.

Note that we have referred consistently to the release of energy
through neutrinos in the flow. This is simply because it is the
physical ingredient we can accurately compute and follow reliably,
serving as a tracer of central engine activity. This does not mean
that neutrinos alone are necessarily capable of powering the observed
emission during flares, as was noted, e.g. for GRB050724 by
\citet{fan05}. Qualitatively, the argument applies to any form of
energy extraction that is dependent upon a mass reservoir, for
example, magnetically driven outflows, with $B^{2}\simeq f \rho c_{\rm
  s}^{2}$ in the disk and $f\leq 1$. In such a scenario, fields
anchored in the disk can drive outflows with corresponding
luminosities reaching $L_{\rm B}\simeq 10^{48}$erg~s$^{-1}$
\citep{bp82}, which are high enough to account for the observed power
outputs.

The total neutrino luminosity including the prompt phase, plotted in
Figure~\ref{fig:ringlight}, shows that the interval between the start
of the evolution and the maximum brightness can reach a minute or more
depending on the intensity of viscous transport. The key ingredient in
the evolution is that the available mass not be driven off the disk
before a substantial fraction of it can accrete onto the BH and
release its gravitational binding energy. The computation of the
cooling rates is crucial in this respect as one needs to maintain an
accretion efficiency of a few per cent in order to successfully
transfer the energy to neutrinos. 

The actual shape of the lightcurve is sensitive to how the energy is
reprocessed into radiation, which in turn depends on the Lorentz
factor of the outflow before and after the mass depletion phase
occurs. This would determine whether internal dissipation takes place
in the secondary outflow before it interacts with previously ejected
material or not \citep{rr01}.

\section{Discussion and conclusions}

The global properties of accretion flows are intimately related to the
cooling mechanisms present (witness the relevance of the two most
widely known limiting solutions, standard Shakura \& Sunayev (SS)
models and Advection Dominated Accretion flows (ADAFs) over the past
35 years). For stationary solutions, the mass and energy fed into the
flow fix the resulting properties. However, when temporal variations
are allowed, transitions from one regime to another can naturally
occur. The solutions exhibited here are well described by the
self-similar temporal decay computed for isolated fluid rings with
efficient cooling by \citet{cannizzo90} and \citet{metzger08}, but also switch to
a different mode, and rapidly decay when the disk mass drops below the
threshold required to sustain adequate cooling. In this particular
instance, the mechanism responsible for mass depletion is a
combination of accretion and neutron and proton recombination into
$\alpha$ particles. This phase transition then effectively allows the
secondary episode of accretion to play an important role with
potentially observable consequences.

We have shown here how the injection of material from tidal tails
formed during the initial disruptive encounter of compact objects can
re-energize the accretion disk provided that the fall-back mass
dominates over the remnant disk at the time when it is re-injected at
its circularization radius. It must be emphasized that the evolution
of this fluid then proceeds not on the relatively short fall-back time
scale itself \citep{rossi09}, but on the viscous time of
the resulting ring. Due to the dynamics of the encounter, the specific
angular momentum can be large enough to increase the characteristic
orbital radius and thus account for longer time scales. Upon return to
the vicinity of the BH, the gas can efficiently cool and accrete. The
resulting luminosity, integrated energy and time scale for delay can
account for the observed activity at late times in SGRBs, provided a
sufficiently efficient mechanism is available to tap the available
energy (e.g., magnetic fields). The details of fall back depend on
many factors \citep{lrr07}, including the neutron star equation of
state, the initial mass ratio of the merging binary, the nature of the
components (double NS binary vs. NS-BH system) and the dynamical
interaction (merger in a binary vs. collision in a dense stellar
environment). All of these would contribute to substantial diversity
in the outcome, and it is possible that many events would have no
observable signal arising from such tails, just as not all mergers may
produce a SGRB.

Late flares have been observed to date in both long and short GRBs
(see e.g., \citet{chincarini07}), and the current proposal applies
only to their production in short events, given the assumptions about
the progenitor.  However \citet{king05} pointed out that fragmentation
of a rapidly rotating core in the collapsar model could in principle
produce secondary episodes of activity after the prompt phase. Thus,
there could be a common underlying cause in both classes in terms of
renewed activity in the central engine through large scale flow
dynamics.

\acknowledgments We thank J. Cannizzo, N. Gehrels, D. Lazzati,
B. Metzger, A. Piro, E. Quataert, M. Rees, L. Roberts, E. Rossi and
S. Rosswog for comments and discussions, as well as the referee for
comments and suggestions on the manuscript. WL is particularly
grateful to A.R. King for his glimpses of the obvious. This work was
supported in part by CONACyT (45845E, WL), PAPIIT-UNAM (IN113007, WL),
the Packard Foundation (ER-R), NASA (Swift NX07AE98G, ER-R) and DOE
SciDAC (DE-FC02-01ER41176, ER-R).  DL-C is supported by a CONACyT
graduate scholarship. WL thanks the Department of Astronomy and
Astrophysics at the University of California, Santa Cruz for
hospitality.

\end{document}